# Merchandise Recommendation for Retail Events with Word Embedding Weighted Tf-idf and Dynamic Query Expansion

Ted Tao Yuan
eBay Inc.
San Jose, California, USA
teyuan@ebay.com

Zezhong Zhang
eBay Inc.
San Jose, California, USA
zezzhang@ebay.com

## ABSTRACT

To recommend relevant merchandises for seasonal retail events, we rely on item retrieval from marketplace inventory. With feedback to expand query scope, we discuss keyword expansion candidate selection using word embedding similarity, and an enhanced tf-idf formula for expanded words in search ranking.

## 1 INTRODUCTION

On eBay, sellers tend to include seasonal event words in listing titles, i.e., jewelry for *Valentine's Day* gift. Such items tend to sell well during the event time period. A retail event model comprises the event's seasonal time window, and its relevance context in terms of keywords. For example, a Valentine's Day retail event is generally from mid-January to mid-February, with curated context keywords like *valentine, jewelry*, etc. In this paper we discuss a way to expand beyond the given context keywords using word embedding, and incorporate query relevance weight in item retrieval ranking for event based merchandise recommendation.

## 2 APPROACH

### 2.1 Item Retrieval And Word Embedding

Given event context keywords, i.e., jewelry for Valentine's Day, we want to retrieve all relevant items sold in the past during the event time period.

Fig.1 shows the process. Item titles (prefixed with category name) as bag of words (BOW), are matched case-insensitively with selected event keywords (partial or all). We rank all retrieved items by the sum of tf-idf scores from matched words, and keep the items with total tf-idf scores above a threshold. The retrieval based system works well to discover relevant merchandises for any event with given seed keywords.

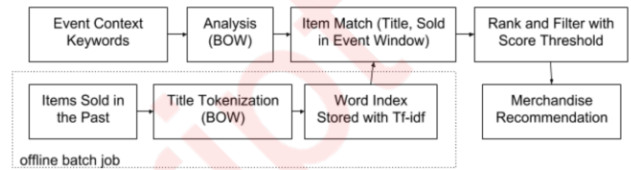

**Figure 1: An event relevant item recommendation system.**

Word embedding [1], is an effective way to generate vector representations of words given a large text corpus. It preserves the linguistic context of words in the item title corpus. Word2vec [2] provides likelihood probability score, or similarity, if two words i and j tend to stay together in the item corpus. The score is bounded and computed with cosine distance,

$$sim(i, j) = cosine\ distance(embed(i), embed(j)) \qquad (1)$$

In our case, it can be used to compute the probability of a word occurring in the neighborhood of given event context keywords.

### 2.2 Expand Event Keyword Scope

One of the issues with the above simple item retrieval based recommendation is that it misses items that relate to an event but the titles do not contain the event seed keywords. For example, some European sellers spell "jewelry" as "jewellery" in their items. More importantly, word meaning changes from season to season that leads to different merchandises, see Table 1 for examples. Word embedding can capture the changing seasonal synonyms of event keywords if we train embedding models on items sold during the months of a season.

**Table 1: Similar Words by Season**

| Word | Month | Similar Words |
|---|---|---|
| valentine | February | cupid, hearts, conversation |
|  | August | happy, birthday, graduation |
| school | April | yearbook, backpacks, college |
|  | August | backpacks, schoolbag, bookbag |

With embedding, we could explore related merchandise in the embedding space using algorithms like word mover distance [3], weighted word embedding aggregation [4], et. al. However, we want more control of the process to separate item retrieval from embedding quality, and combine them via query expansion.

## 3 DYNAMIC QUERY EXPANSION WITH EMBEDDING AND WEIGHTED TF-IDF

As discussed in the Probabilistic Relevance Framework (PRF) [5], query expansion arises from relevance feedback. In our item recommendation case we want to include additional words with most similar meanings during the event time window.

In PRF, the offer weight of a query expansion word candidate is proportional to the word's document weight and its relevance to the query intent. The offer weight was used to rank and select query expansion candidates. The selected words along with the original query are included in normal retrieval and ranking.

In our case, event seed keywords define the query intent. We want other highly related words to be included in the retrieval and ranking, but controlled by the word correlation weight. Using same notation as in [5], i.e., $\propto_q$ means rank equivalence, the probability of relevance of document (item title) d and query q is

$$P(rel \mid d, q) \propto_q \frac{P(d|rel, q)}{P(d|\overline{rel}, q)} \approx \prod_{i \in V} \frac{P(TF_i = tf_i \mid rel, q)}{P(TF_i = tf_i \mid \overline{rel}, q)} \quad (2)$$

$$\approx \prod_{i \in q} \frac{P(TF_i = tf_i|rel, q)}{P(TF_i = tf_i|\overline{rel}, q)} \times \prod_{j \in q'} \frac{P(TF_j = tf_j|rel, q)}{P(TF_j = tf_j|\overline{rel}, q)} \quad (3)$$

$$\propto_q \sum_{i \in q} \log \frac{P(TF_i = tf_i|rel, q)}{P(TF_i = tf_i|\overline{rel}, q)}$$
$$+ \sum_{j \in q'} \log \frac{P(TF_j = tf_j|rel, q)}{P(TF_j = tf_j|\overline{rel}, q)} \quad (4)$$

$$= \sum_{i \in q} U_i(tf_i) + \sum_{j \in q'} U'_j(tf_j, q) \quad (5)$$

where V is vocabulary, q' are expanded words in V but not in q. Given a non-stop-word i in q, candidates for q' are selected as
1. compute similarity score (> 0.6) to i from Eq. (1),
2. rank by the score and select the top k (<5) words.

As in PRF, in Step (2), we assume the conditional probability of each word can be evaluated independently, and expand the document probability as product over the words of the vocabulary (Naïve Bayes). Note we keep q' in Step (3) and use monotonic transformation in Step (4).

The q' related second term in Step (5) may be estimated heuristically through q with a parameter $S(j, q)$ for $j \in q'$,
$$U'_j(tf_j, q) \approx U_j(tf_j) \times S(j, q)$$

where $S(j, q)$ measures relevance of word j to query intent q, and $S(j, q) \approx \max(sim(j, m))$ with m ∈ (non-stop-words in q).

If we use tf-idf as an approximation of U(tf), and combine all words in q and q', we have the following formula for ranking,

$$P(rel \mid d, q) \propto_q \sum_{i \in q \cup q'} tf(d, i) \times idf(i) \times \delta(i, q) \quad (6)$$

where $\delta(i, q) = 1$ if $i \in q$; and $\delta(i, q) = S(i, q)$ if $i \in q'$. We call it word embedding weighted tf-idf for query expanded words.

## 4 EXPERIMENTS AND RESULTS

eBay has vast and diverse online merchandises. It is desirable to classify inventory in the online marketplace by year round seasonal and social events. We built our item retrieval system on Hadoop/Spark platform that allows us to access billions of items during past seasons. We segment sold items by month, and train word embedding model for each month to capture seasonal word synonyms. For a retail event, we search relevant items by matching expanded event keywords with item titles, and rank the items with Formula (6). Items with scores above a threshold are selected for further recommendation processing.

In our case of given event context seed keywords, we focus more on expanding the inventory coverage for an event. Albeit subjective in recall judgement, by systematically adding strongly correlated words that are selected through seasonal word embedding similarity, we can gradually increase relevant inventory scope for recommendation, see Table 2 for examples.

**Table 2: Expansion Examples and Recall Impact**

| Seed Query | Expanded Word : Weight | Recall % Increase (US site) |
|---|---|---|
| valentines day jewelry | jewellery : 0.93 | < 1% |
| back to school | bookbag : 0.8; backpacks : 0.7 | > 200% |

## 5 CONCLUSION

For seasonal event recommendation, we create monthly word embedding models based on item titles. We discuss a way to use word similarity scores from the word embedding models to rank and select query expansion candidates with given retail event seed keywords. Further we enhance tf-idf ranking scores of the expanded words using the embedding similarity as query relevance weights. We believe it can be applicable to other document weight based ranking as well, i.e., replace tf-idf with BM25. The embedding enhanced retrieval and ranking scheme helps us discover more relevant merchandises systematically for event based recommendation.